\documentclass[sigplan,screen]{acmart}
\usepackage{xspace}
\usepackage{multirow}
\usepackage{microtype}
\usepackage{color, colortbl}
\definecolor{RowGray}{gray}{0.97}

\newcommand{\Copilot}{GitHub Copilot\xspace}
\newcommand{\GitHubCopilot}{OurTool\xspace}
\newcommand{\light}[1]{\textcolor{lightgray}{#1}}
 
\AtBeginDocument{%
  \providecommand\BibTeX{{%
    \normalfont B\kern-0.5em{\scshape i\kern-0.25em b}\kern-0.8em\TeX}}}

\setcopyright{rightsretained}
\acmPrice{}
\acmDOI{10.1145/3520312.3534864}
\acmYear{2022}
\copyrightyear{2022}
\acmSubmissionID{pldiws22mapsmain-p3-p}
\acmISBN{978-1-4503-9273-0/22/06}
\acmConference[MAPS '22]{Proceedings of the 6th ACM SIGPLAN International Symposium on Machine Programming}{June 13, 2022}{San Diego, CA, USA}
\acmBooktitle{Proceedings of the 6th ACM SIGPLAN International Symposium on Machine Programming (MAPS '22), June 13, 2022, San Diego, CA, USA}
\begin{document}

\title{Productivity Assessment of Neural Code Completion}

\author{Albert Ziegler, Eirini Kalliamvakou, Shawn Simister, Ganesh Sittampalam, Alice Li, Andrew Rice, Devon Rifkin, and Edward Aftandilian}
\email{{wunderalbert, ikaliam, narphorium, hsenag, xalili, acr31, drifkin, eaftan}@github.com}
\affiliation{%
  \institution{GitHub, Inc.}
  \country{USA}
}

\renewcommand{\shortauthors}{Ziegler, Kalliamvakou, Simister, Sittampalam, Li, Rice, Rifkin, and Aftandilian}

\begin{abstract}
  Neural code synthesis has reached a point where snippet generation is accurate enough to be considered for integration into human software development workflows. Commercial products aim to increase programmers’ productivity, without being able to measure it directly. In this case study, we asked users of \Copilot about its impact on their productivity, and sought to find a reflection of their perception in directly measurable user data. We find that the rate with which shown suggestions are accepted, rather than more specific metrics regarding the persistence of completions in the code over time, drives developers’ perception of productivity.
\end{abstract}





%% turn on page numbers
\settopmatter{printacmref=true}

\maketitle

\section{Introduction}
\label{sec:Intro}

Code completion systems that offer suggestions to a developer on the basis of contextual information from the IDE have been shown to be by far the most frequently used kind of programmer assistance~\cite{DBLP:conf/wcre/AmannPNM16}. One common example is that of proposing a list of method names based on the type of a variable. Neural code synthesis approaches to code completion generate suggestions by using a language model to predict what the user might type next (the completion) from the context of what they are working on at the moment (the prompt)~\cite{DBLP:journals/corr/abs-2108-07732}.  Rather than focusing on a particular task (such as suggesting a method to call), neural code synthesis predicts arbitrary sections of code, and rather than generating single tokens, these systems might predict multiple lines of code at once.

The potential benefits of generating large sections of code automatically are huge, but evaluating these systems is challenging. Offline evaluation where the system is shown a snippet of code with (say) an identifier removed and then asked to complete it is difficult not least because for longer completions there are many acceptable alternatives and no straightforward mechanism for labeling them automatically~\cite{DBLP:journals/corr/abs-2107-03374}. An additional step taken by some researchers~\cite{DBLP:conf/msr/SvyatkovskiyLHR21, DBLP:journals/corr/abs-2105-05991, DBLP:conf/icse/AyeKL21} is to use online evaluation and track the frequency of real users accepting suggestions, assuming that the more contributions a system makes to the developer’s code, the higher its benefit. The validity of this assumption is not obvious when considering issues such as whether two short completions are more valuable than one long one or other human factors such as whether reviewing suggestions is detrimental to programming flow.

Neural synthesis tools such as \Copilot\footnote{\url{https://copilot.github.com/}}, Kite\footnote{\url{https://www.kite.com}}, and TabNine\footnote{\url{https://tabnine.com}} suggest code snippets within an IDE with the explicitly stated intention to increase a user's productivity. Developer productivity has many aspects, and a recent study has shown that tools like these are helpful in ways that are only partially reflected by measures like completion times for standardized tasks \cite{10.1145/3491101.3519665}. Alternatively, we can leverage the developers themselves as expert assessors of their own productivity. This meshes well with current thinking in software engineering research which suggests measuring productivity on multiple dimensions and using self-reported data~\cite{forsgren2021space}. We will thus focus on studying \emph{perceived} productivity. 

In this paper we investigate whether usage measurements of developer interactions with \Copilot can be used to predict perceived productivity as reported by developers.  We analyze $2{,}631$ survey responses from developers using \Copilot and match their responses to usage measurements collected from the IDE.  We consider acceptance counts and more detailed measures of contribution such as the amount of code contributed by \Copilot, and the rate of acceptances which subsequently persist in the code unchanged. \textbf{We find that acceptance rate of shown suggestions is a better predictor of perceived productivity than the alternative measures}. We also find that acceptance rate varies significantly over our developer population as well as over time, and present a deeper dive into some of these variations.

Our results support the principle that acceptance rate can be used for coarse-grained monitoring of the performance of a neural code synthesis system. In particular, the ratio of shown suggestions being accepted correlates better than more detailed measures of contribution. However, other approaches remain necessary for fine-grained investigation due to the many human factors involved.

\section{Background}
\label{sec:Background}

Offline evaluation of code completion can have shortcomings even in tractable circumstances where completions can be labeled for correctness. For example, a study of $15{,}000$ completions by $66$ developers in Visual Studio found significant differences between synthetic benchmarks used for model evaluation and real-world usage~\cite{DBLP:conf/icse/HellendoornPGB19}. The evaluation of context-aware API completion for Visual Studio IntelliCode considered Recall@5---the proportion of completions for which the correct method call was in the top 5 suggestions. This metric fell from $90\%$ in offline evaluation to $70\%$ when used online~\cite{DBLP:conf/msr/SvyatkovskiyLHR21}.

Due to the diversity of potential solutions to a multi-line completion task, researchers have used software testing to evaluate the behaviour of completions. Competitive programming sites have been used as a source of such data~\cite{DBLP:conf/nips/KulalPC0PAL19,DBLP:journals/corr/abs-2105-09938}  as well as hand-written programming problems~\cite{DBLP:journals/corr/abs-2107-03374}. Human-written tests can be augmented with automatically generated tests to reduce false-positive rates in which erroneous programs are accepted~\cite{alphacode}. Online evaluation remains important for general code completion tools because one needs to understand how well performance on programming competition data generalizes to interactive development in an IDE.

In this work we define acceptance rate as the fraction of completions shown to the developer that are subsequently accepted for inclusion in the source file.  The IntelliCode Compose system uses the term CTR (Click Through Rate) for this and reports a value of $10\%$ in online trials~\cite{DBLP:conf/sigsoft/SvyatkovskiyDFS20}. An alternative measure is that of DCPU (Daily Completions accepted Per User) for which a value of around $20$ has been reported~\cite{DBLP:journals/corr/abs-2105-05991,DBLP:conf/icse/AyeKL21}. To calculate acceptance rate one must, of course, normalize DCPU by the time spent coding each day.  For context, in our study \Copilot has an acceptance rate of $27\%$ and a mean DCPU in excess of $31$. These differences are presumably due to differences in the kinds of completion offered, or perhaps to user interface choices. We discuss later how developer objectives, choice of programming language and even time of day seem to affect our data. Such discrepancies highlight the difficulty in using acceptance rate to understand the value of a system.

There is some evidence that acceptance rate (and indeed correctness) might not tell the whole story. One survey of developers considered the use of AI to support translation between programming languages and found indications that developers tolerated, and in some cases valued, erroneous suggestions from the model~\cite{DBLP:conf/iui/WeiszMHRRMAT21}. 

Measuring developer productivity through activity counts over time (a typical definition of productivity borrowed from economics) disregards the complexity of software development as they account for only a subset of developer outputs. A more holistic picture is formed by measuring \textit{perceived} productivity through self-reported data across various dimensions~\cite{forsgren2021space}, and supplementing it with automatically measured data~\cite{beller2020mind}. In our investigation we used the SPACE framework~\cite{forsgren2021space} to design a survey that captures self-reported productivity, and paired the self-reported data with usage telemetry.     

To the best of our knowledge, this is the first study of code suggestion tools establishing a clear link between usage measurements and developer productivity or happiness. A previous study comparing GitHub Copilot against IntelliCode with 25 participants found no significant correlation between task completion times and survey responses~\cite{evalcopilot}. Another study considered the benefits of using a plugin converting natural language prompts to code~\cite{DBLP:journals/corr/abs-2101-11149}. It found no statistically significant improvements in task completion time or task correctness despite positive qualitative survey results (possibly due to small sample size).

\section{Data and Methodology}
\label{sec:Methodology}

\subsection{Usage Measurements}
\label{sec:UsageMeasurements}
\Copilot provides code completions using OpenAI Codex \cite{DBLP:journals/corr/abs-2107-03374} , which is a version of GPT-3 that has been tuned on publicly available source code. It runs within the IDE and at appropriate points sends a completion request to a cloud-hosted instance of the neural model. Completion requests contain a prompt drawn from the code currently in the IDE.  \Copilot can generate completions at arbitrary points in code rather than (say) only being triggered when a developer types a period for invoking a method on an object. We use a variety of rules to determine appropriate points to request a completion; to abandon requests if the developer has moved on before the model is ready with a completion; and to determine how much of the response from the model to surface as a completion.

\begin{figure}
  \includegraphics[width=0.48\textwidth]{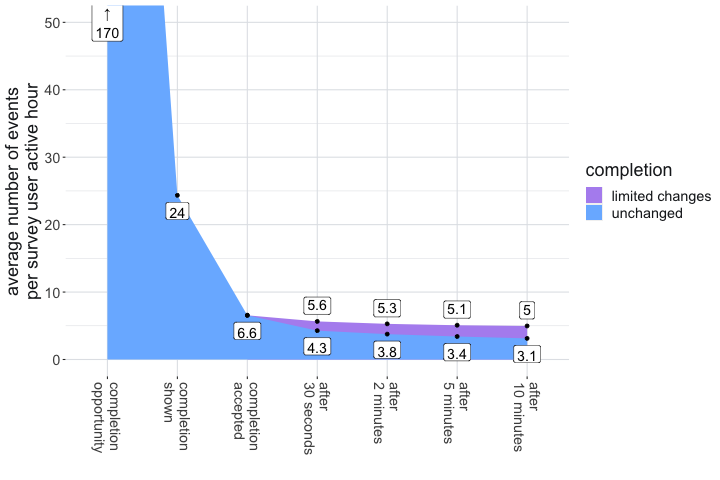}
  \caption{\Copilot's code completion funnel.}
  \label{fig:metric_distribution}
\end{figure}

\begin{table*}
    \caption{Developer usage events collected by \Copilot.}
    \label{tab:eventscollected}
    \begin{tabular}{p{3.4cm}|p{13.4cm}}
 \hline
  \texttt{opportunity} & a heuristic-based determination by the IDE and the plugin that a completion might be appropriate at this point in the code (e.g. the cursor is not in the middle of a word) \\
\rowcolor{RowGray}  \texttt{shown} & completion shown to the developer  \\

  \texttt{accepted} & completion accepted by the developer for inclusion in the source file \\
\rowcolor{RowGray}  \texttt{accepted\_char} &  the number of characters in an accepted completion  \\
  \texttt{mostly\_unchanged\_X} & completion persisting in source code with limited modifications (Levenshtein distance less than 33\%) after X seconds, where we consider a duration of 30, 120, 300, and 600 seconds \\
\rowcolor{RowGray}\texttt{unchanged\_X} & completion persisting in source code unmodified after X seconds. \\
  (active) \texttt{hour} & an hour during which the developer was using their IDE with the plugin active \\
      \hline

    \end{tabular}
\end{table*}

We make usage measurements for each developer by counting the events shown in Table~\ref{tab:eventscollected}, collected for all users of \Copilot according to our terms of usage.\footnote{\url{https://docs.github.com/en/github/copilot/github-copilot-telemetry-terms}}

Our measures of persistence go further than existing work which stops at acceptance.  The intuition here is that a completion which is accepted into the source file but then subsequently turns out to be incorrect con be considered to have wasted developer time both in reviewing it and then having to go back and delete it again.  We also record \emph{mostly} unchanged completions, reasoning that a large completion requiring a few edits might still be a positive contribution. It is not clear how long after acceptance one should confirm persistence and so we consider a range of options. 

The events pertaining to completions form a funnel which we show quantitatively in Table \ref{fig:metric_distribution}. We include a summary of all data in Appendix~\ref{appendix:usage}.

We normalize these measures against each other and write \texttt{X\_per\_Y} to indicate we have normalized metric \texttt{X} by metric \texttt{Y}. For example: \texttt{accepted\_per\_hour} is calculated as the total number of \texttt{accepted} events divided by the total number of (active) \texttt{hour} events.

\begin{table*}
  \caption{The core set of measurements considered in this paper\label{tab:coreset}.}
    \begin{tabular}{p{3.4cm}|p{6.6cm}|p{5.6cm}}
          \hline
          \textbf{Natural name} & \textbf{Explanation} & \textbf{Definition} \\
          \hline
Shown rate & Percentage of completion opportunities that resulted in a completion being shown to the user& \texttt{shown\_per\_opportunity}\\

\rowcolor{RowGray}Acceptance rate & Percentage of shown completions accepted by the user &\texttt{accepted\_per\_shown}\\

Persistence rate & Percentage of accepted completions unchanged after 30, 120, 300, and 600 seconds & \texttt{unchanged\_X\_per\_accepted}\\

\rowcolor{RowGray}Fuzzy persistence rate & Percentage of accepted completions mostly unchanged after 30, 120, 300, and 600 seconds & \texttt{mostly\_unchanged\_X\_per\_accepted}\\

Efficiency &  Percentage of completion opportunities that resulted in a completion accepted and unchanged after 30, 120, 300, and 600 seconds & \texttt{accepted\_X\_per\_opportunity}, \texttt{unchanged\_X\_per\_opportunity}\\

\rowcolor{RowGray}Contribution speed & Number of characters in accepted completions per distinct, active hour & \texttt{accepted\_char\_per\_hour} \\

Acceptance frequency &  Number of accepted completions per distinct, active hour & \texttt{accepted\_per\_hour}\\

\rowcolor{RowGray}Persistence frequency &  Number of unchanged completions per distinct, active hour & \texttt{unchanged\_X\_per\_hour}\\

Total Volume &  Total number of completions shown to the user & \texttt{shown}\\

\rowcolor{RowGray}Loquaciousness &  Number of shown completions per distinct, active hour &\texttt{shown\_per\_hour}\\

Eagerness &  Number of shown completions per opportunity &\texttt{shown\_per\_opportunity}\\
      \hline
  \end{tabular}
\end{table*}

Table~\ref{tab:coreset} defines the a core set of metrics which we feel have a natural interpretation in this context. We note that there are other alternatives here and we incorporate these in our discussion where relevant.

\subsection{Productivity Survey}

To understand users’ experience with \Copilot, we emailed $17{,}420$ users providing them with a link to complete an online survey. These were participants of the unpaid technical preview using \Copilot with their everyday programming tasks. The only selection criterion was having previously opted in to receive communications. Between 10$^{\mathrm{th}}$ February 2022 and 6$^{\mathrm{th}}$ March 2022, we received $2{,}047$ responses we could match to usage measurements during the 4 week period leading up to February 12$^{\mathrm{th}}$ March 2022. We focus on usage data from this period, since the vast majority ($>80\%$) of survey users had filled out their survey by then.

\begin{figure}
  \includegraphics[width=0.48\textwidth]{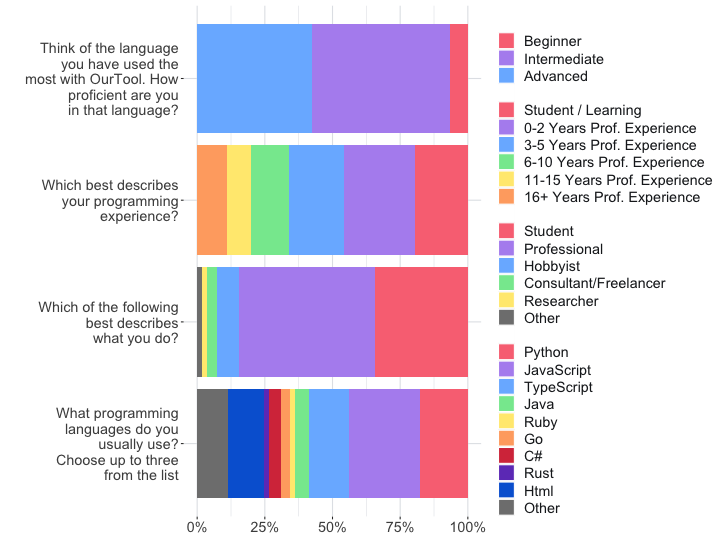}
  \caption{Demographic composition of survey respondents.}
  \label{fig:demographics}
\end{figure}

The survey questions contained multiple choice questions, in particular regarding demographic information (shown in Figure~\ref{fig:demographics}) and Likert-style questions about different aspects of productivity, which were randomized in their order of appearance to the user. Figure~\ref{fig:demographics} shows the demographic composition of our respondents. We note the significant proportion of professional programmers who responded.

The SPACE framework~\cite{forsgren2021space} defines 5 dimensions of productivity:  \textbf{S}atisfaction and well-being, \textbf{P}erformance, \textbf{A}ctivity, \textbf{C}ommunication and Collaboration, \textbf{E}fficiency and Flow. We use 4 of these (S,P,C,E) since self reporting on (A) is generally considered inferior to direct measurement. We included $11$ statements covering these $4$ dimensions in addition to a single statement ``I am more productive when using \Copilot''. For each self-reported productivity measure, we encoded its five ordinal response values to numeric labels (1 = Strongly Disagree, $\ldots$, 5 = Strongly Agree).  We include the full list of questions and their coding to the SPACE framework in Appendix~\ref{appendix:surveyaspects}.

Early in our analysis we found that the usage metrics we describe in Section \ref{sec:UsageMeasurements} below corresponded similarly to each of the measured dimensions of productivity, and in turn these dimensions were highly correlated to each other (for details see Figure \ref{fig:correlation} and Appendix~\ref{appendix:correlation}). We therefore added an aggregate productivity score calculated as the mean of all 12 individual measures (excluding skipped questions). This average can only serve as a rough proxy for the much more complex concept of productivity, but facilitates recognition of overall trends, which may be less discernible on individual variables due to higher statistical variation.  

For reproducibility and transparency, the full data set of these aggregate productivity scores together with the usage measurements considered in this article is available at \url{https://github.com/wunderalbert/prod-neural-materials}.

\section{What Drives Perceived Productivity?}
\label{sec:Productivity}

\begin{figure*}
  \includegraphics[width=0.98\textwidth]{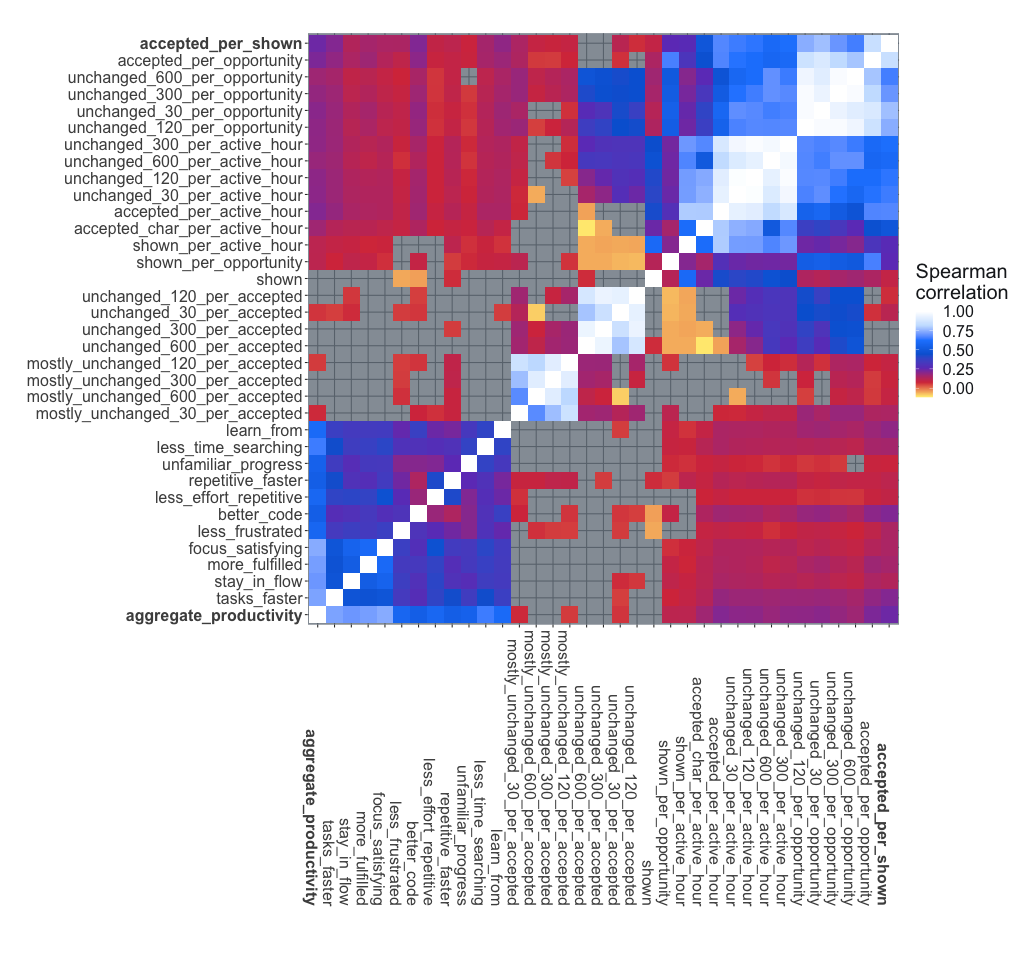}
  \caption{Correlation between metrics.}
  \small
  Metrics are ordered by similarity based on distance in the correlation matrix, except for manually fixing the aggregate productivity and acceptance rate at the end for visibility.
  \label{fig:correlation}
\end{figure*}

To examine the relationship between objective measurements of user behavior and self-reported perceptions of productivity, we used our set of core usage measurements (Table~\ref{tab:coreset}). We then calculated Pearson’s R correlation coefficient and the corresponding p-value of the F-statistic between each pair of usage measurement and perceived productivity metric. Next, we computed a PLS regression from all usage measurements jointly. Finally, we perform incremental feature selection by analyzing the significance of a univariate model where each usage measurement seeks to predict the residuals of a model fit with varying numbers of other metrics; this allows us to more directly rank each metric.

We summarize these results in Figure~\ref{fig:correlation} showing the correlation coefficients between all measures and survey questions. The full table of all results is included in Appendix~\ref{appendix:correlation}.

\textbf{Across all three analyses, we find that acceptance rate (\texttt{accepted\_per\_shown}) most positively predicts users’ perception of productivity, although, given the confounding and human factors, there is still notable unexplained variance.}

%\subsection{Correlations}

Of all usage measurements, acceptance rate correlates best with aggregate productivity ($\rho=0.24$, $P<0.0001$). This measurement is also the best performing for at least one survey question in each of the SPACE dimensions. This correlation is high confidence but leaves considerable unexplained variance. Below we explore improvements from combining multiple usage measurements together.

Looking at the more detailed metrics around persistence, we see that persistence over shorter time periods is generally better than over longer periods. This is intuitive in the sense that shorter periods move the measure closer to acceptance rate.  We also expect that at some point after accepting the completion it becomes simply part of the code and so any changes (or not) after that point will not be attributed to \Copilot.  All persistence measures were less well correlated than acceptance rate.

In order to assess the different metrics in a single model in a way that is still robust against their strong collinearity and unaffected by the decision whether or not to include highly similar metrics, we ran a regression using projection on latent structures (PLS), which captures the common variation of these variables as is linearly connected to the aggregate productivity~\cite{WOLD2001109}. The first component, to which every metric under consideration contributes positively, explains $43.2\%$ of the variance. The second component captures the acceptance rate / change rate dichotomy; it explains a further $13.1\%$. 

\begin{figure}
  \includegraphics[width=0.48\textwidth]{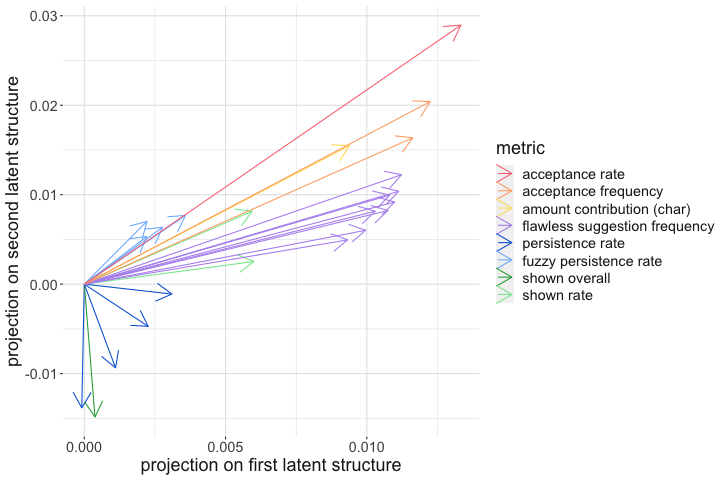}
  \caption{Different metrics clustering in latent structures predicting perceived productivity. We color the following groups: \emph{flawless suggestions} (anything counting the number of unchanged suggestions), \emph{persistence rate} (ratio of accepted suggestions that are unchanged), and \emph{fuzzy persistence rate} (ratio of accepted suggestions that are mostly unchanged).}
  \label{fig:clustering}
\end{figure}

The results of both the individual correlations as well as the PLS strongly point to acceptance rate being the most immediate indicator of perceived productivity.

But what about combinations of metrics? We aim to quantify the extra information provided by one metric over a set of others. In the vein of incremental feature selection, we fit an additional predictor to the residual of a model represented by already selected metrics. Starting from the best single predictors, Table~\ref{table:incremental} shows the next most useful predictors that predict the residual of the acceptance rate model at $\mathrm{p} < 0.05$. Given a model fit to acceptance rate, adding the shown frequency or rate, as well as either amount of accepted characters or accepted completions per hour each further improve predictive ability at statistically significant levels. No other additions were statistically significant for further iterations.

\begin{table}
\caption{Incremental benefit of additional metrics.}
\label{table:incremental}
\centering
\begin{tabular}[t]{lrrr}
\toprule
& \textbf{p-value} & \textbf{univ. coef.}\\
\midrule
{\tt accepted\_per\_shown }&<0.0001&0.13\\
\rowcolor{RowGray}\hspace{0.5cm}+ {\tt shown\_per\_hour }&0.04&+0.03\\
\hspace{0.5cm}+ {\tt accepted\_char\_per\_hour }&0.04&+0.03\\
\rowcolor{RowGray}\hspace{0.5cm}+ {\tt shown\_per\_opportunity }&0.04&+0.03\\
\hspace{0.5cm}+ {\tt accepted\_per\_hour }&0.05&+0.02\\
\hline
\rowcolor{RowGray}{\tt accepted\_per\_opportunity }&<0.0001&0.12\\
\hspace{0.5cm}+ {\tt accepted\_char\_per\_hour }&<0.0001&+0.04\\
\rowcolor{RowGray}\hspace{0.5cm}+ {\tt accepted\_per\_hour }&0.01&+0.03\\
\hspace{0.5cm}+ {\tt unchanged\_30\_per\_hour }&0.01&+0.03\\
\rowcolor{RowGray}\hspace{0.5cm}+ {\tt accepted\_per\_shown }&0.02&+0.03\\
\bottomrule
\end{tabular}
\end{table}%

So even if acceptance rate may be the best of the metrics we considered, it is beneficial to combine with others to get a fuller picture.
\section{What Drives Acceptance Rate?}
\label{sec:Acceptance}

\subsection{Language Use}

We are aware that there are significant differences for how \Copilot performs for different programming languages. The most common languages among our user base are TypeScript ($24.7\%$ of all shown completions in the observed time frame, $21.9\%$ for users in survey), JavaScript ($21.3\%$, $24.2\%$), and Python ($14.1\%$, $14.5\%$). The latter two enjoy higher acceptance rates, possibly hinting at a relative strength of neural tooling versus deductive tooling for untyped languages. Regardless of language, survey participants had a slightly higher acceptance rate than the whole user base.

\begin{figure}[h]
  \includegraphics[width=0.48\textwidth]{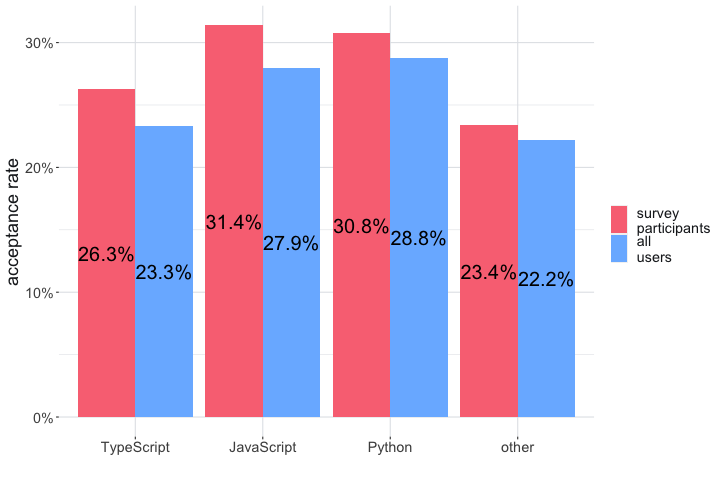}
  \caption{Programming language use by survey participants vs. all users.}
  \label{fig:languages}
\end{figure}

This difference in language acceptance can not explain away the effects from Section~\ref{sec:Productivity}: when considering the linear regression for perceived productivity from acceptance rate, only $17\%$ of the explained variance can be attributed to language. On the other hand, however, $38\%$ of the variance of the different levels for the languages' average perceived productivity can be explained by the acceptance rates in Figure~\ref{fig:languages} above (using a linear model factoring through acceptance rate).

\subsection{Circadian and Weekly Rhythms}

For coherence in the meaning of timestamps and weekdays, all data in this section was restricted to users from the United States (whether in the survey or not). We used the same time frame as for the investigation in Section~\ref{sec:Productivity}.

\begin{figure}
  \includegraphics[width=0.48\textwidth]{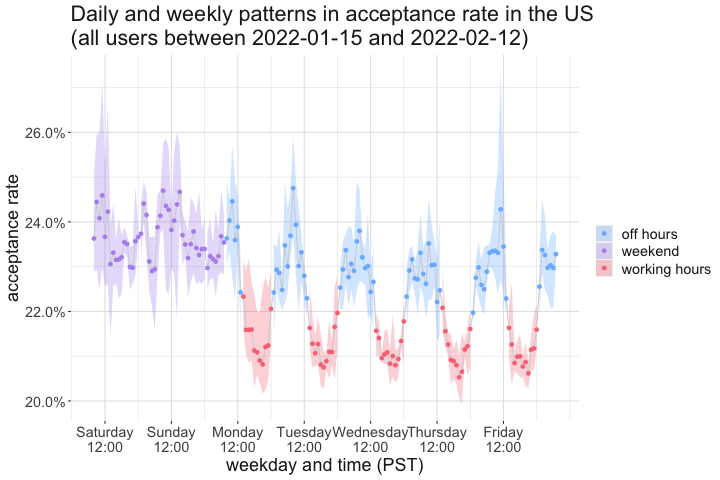}
  \caption{Average acceptance rate for hour-long time buckets during the week. Each point represents the average for such a bucket, whereas the shaded ribbon represents the min-max variation for single hours during the observed 4 week period.}
  \label{fig:acceptance}
\end{figure}

We observe strong regular patterns in overall acceptance rate (Figure \ref{fig:acceptance}). These lead us to distinguish three different time regimes, all of which are statistically significantly distinct at $\mathrm{p}<0.001\%$ (using a bootstrap test re-sampling the proposed time regimes): 

\begin{itemize}
    \item The weekend: Saturdays and Sundays (day boundaries taken from PST), where the average acceptance rate is comparatively high at $23.5\%$.
    \item Typical non-working hours during the week: evenings after 4 pm PST until mornings 7 am PST, where the average acceptance rate is also rather high at $23\%$.
    \item Typical working hours during the week from 7 am PST to 4 pm PST, where the average acceptance rate is much lower at $21.2\%$.
\end{itemize}

The border between the second and third regime is fuzzy, probably partially due to time zone variation within the United States.

Users’ inclination to accept more suggestions outside of standard working hours could be attributed either to regular changes in the users’ behavior (e.g. accepting more solutions because they are more relaxed), or to changes in the underlying distribution of who is coding and what they are working on (e.g. personal projects being easier to suggest code for).

To distinguish between these two explanations, we trained a model to predict a user’s acceptance rate for a particular time bucket from their usual contribution times (Table \ref{table:circular}). Unexpectedly to us, we found that the actual time bucket mattered very little -- but what did matter was whether it lay in the user’s \textit{usual} time regime. That means a user normally active only during the week accepts fewer solutions on the rare occasions they do code on a weekend, and a user whose activity is normally restricted to working hours accepts fewer solutions when they do venture outside that time range.

\begin{figure}
  \includegraphics[width=0.48\textwidth]{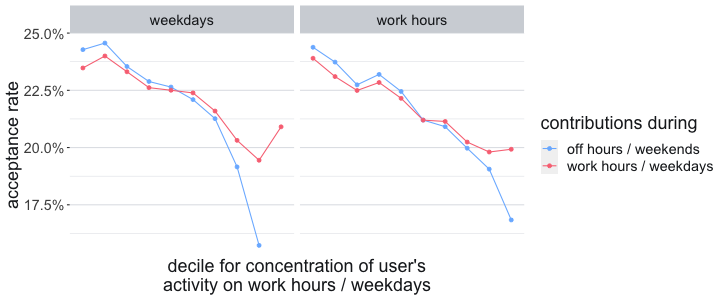}
  \caption{Acceptance rate depending on whether the user is mostly active on weekdays / typical work hours (x-axis), and whether it is actually a weekday / typical office hour (color).}
  \label{fig:workhours}
\end{figure}

\begin{table*}
\caption{Acceptance rate depending on time factors.}
\resizebox{\textwidth}{!}{\begin{tabular}[t]{lrrr|lrrr}
\toprule
& \textbf{coeff} & \textbf{p-value} & \textbf{t-value}& & \textbf{coeff} & \textbf{p-value} & \textbf{t-value}\\
\midrule
\rowcolor{RowGray}weekend user & -0.035 & <0.001 & -24.5 & work hour user & -0.036 & <0.001 & -22.3   \\
on a weekday & -0.004 & 0.004 & -2.9 & during work hours & -0.001 & 0.578 & -0.6  \\
\rowcolor{RowGray}usual day for user & 0.021 & <0.001 & 11.3 & usual time for user & 0.019 & <0.001 & 11.5   \\
\bottomrule
\end{tabular}}
\label{table:circular}
\small
Results of a linear regression of acceptance rate from: 1. user’s percentile value for of how much their activity is concentrated on weekdays (during typical work hours), 2. a categorical variable describing whether the suggestion is actually made on a weekday (during typical work hours), and 3. the proportion of this user’s contributions where that categorical variable would have the same value.
\end{table*}
\section{Threats To Validity}

The principal challenge to this work is that we have only been able to investigate correlation and not causation. We hope to have mitigated this to some extent by selecting ``sensible'' metrics that we can justifiably believe could capture a causal relationship with productivity. Nor do we claim that these metrics themselves directly impact productivity (a developer with a faulty tab-key that accidentally accepts $80\%$ of suggestions without user intention will probably not be extra productive because of it), but only that these metrics are a good \emph{indicator} of an underlying quality that predicts productivity.

We did not set out to accurately predict a survey participant's answers, but merely to find a signal between developers' perceived productivity and usage metrics. Still, we must highlight that our best performing measurement \texttt{acceptance\_per\_shown} has a Pearson coefficient of $0.24$ and so there is a considerable amount of unexplained variance remaining.

User-perceived productivity is also not necessarily actual productivity: seeking to maximise \texttt{acceptance\_per\_shown} might satisfy an individual developer without directly reducing the amount of time it takes them to solve a task. And indeed one study looking at the benefits of GitHub Copilot over IntelliCode found no measurable impact on task completion time despite notably positive feedback from developers~\cite{evalcopilot}. On the other hand, one drawback of task-oriented studies is of how representative the chosen tasks are of real workloads whereas online studies (such as ours) capture authentic activity.

Another substantial caveat is that we only considered a single completion system, in particular with a single fixed neural engine. Alternatives could be different in many aspects that could affect developer attitudes. Factors might include average quality and the latency of completions, their length and even the user interface used to present them.
\section{Conclusion}

Neural code completion systems have the potential to hugely improve developer productivity through their ability to assimilate contextual information about the developer's current activity and then generate substantial completions in response. In this paper we investigated ways of connecting the productivity benefit of \Copilot to usage measurements from developer activity.  Our approach was to seek correlations between our measurements and user-reported productivity from survey results.

In common with prior work we collected measurements about the acceptance of completions, but we also developed measures of persistence. This was based on the idea that for longer completions a developer might have to take more action after accepting a completion such as deleting or correcting an erroneous one. 

We were surprised to find that acceptance rate (number of acceptances normalized by the number of shown completions) was better correlated with reported productivity than our measures of persistence. 

But in hindsight, this makes sense. Coding is not typing, and \Copilot{}'s central value lies not in being the way the user enters the highest possible number of lines of code. Instead, it lies in helping the user to make the best progress towards their goals. A suggestion that serves as a useful template to tinker with may be as good or better than a perfectly correct (but obvious) line of code that only saves the user a few keystrokes.

This suggests that a narrow focus on the correctness of suggestions would not tell the whole story for these kinds of tooling. Instead one could view code suggestions inside an IDE to be more akin to a conversation with a chatbot. We see anecdotal evidence of this in comments posted about \Copilot online (see Appendix~\ref{appendix:comments} for examples) in which users talk about sequences of interactions. A conversation turn in this context consists of the prompt in the completion request and the reply as the completion itself. The developer's response to the completion arises from the subsequent changes which are incorporated in the next prompt to the model. And there are clear programming parallels to factors such as specificity and repetition that have been identified to affect human judgements of conversation quality~\cite{DBLP:conf/naacl/SeeRKW19}. Researchers have already investigated the benefits of natural language feedback to guide program synthesis~\cite{DBLP:journals/corr/abs-2108-07732} and so ours is not a radical proposal. But neither is it one we have seen followed.

In future work, we wish to further explore this analogy, borrowing ideas~\cite{DBLP:conf/inlg/LeeGMWK19} from the evaluation of chatbots and natural language text generation. 
\section{Broader Impact}

A detailed impact analysis of the model that underlies \Copilot may be found in the Appendix of \cite{DBLP:journals/corr/abs-2107-03374}.  In this section, we focus more specifically on the potential impact of using the metrics we have described in this paper to evaluate the success of neural code completion systems.

First, focusing on a single top-level metric such as acceptance rate may bias a tool toward the most popular use cases --- the most popular programming languages, natural languages, IDEs, locations, etc.  Users in underrepresented groups may see lower quality results.  We can mitigate this by slicing our data along the lines described above, and avoiding shipping changes that improve the top-level metric but degrade performance for other slices of the data.

Second, to compute these metrics, we must collect telemetry from users.  Collecting these metrics exposes users to potential security and privacy concerns.  We mitigate this by enacting strict access controls to user data and collaborating with organization and industry experts at protecting user data.

Third, blindly optimizing for a proxy (acceptance rate) for a desired property (usefulness) encourages artificial changes that improve \textit{only} that proxy. For example, cutting code suggestions into half and suggesting both parts consecutively would likely transform one accepted suggestion into two, while not substantially increasing the number of rejections. Thus it would likely increase acceptance rate without substantially increasing, and maybe even while decreasing, user benefit. We can thus not recommend acceptance rate as singular and ultimate criterion of quality -- it will be useful for many applications, e.g. comparing incremental changes to the code generating model, but its validity is limited in other cases, especially those involving significantly changed operational parameters.

%\afterpage{\clearpage}

\section*{Acknowledgments}

We thank the GitHub Copilot team for their help, and in particular Krzysztof Cieslak and Johan Rosenkilde for implementing the highly complex telemetry of suggestion fate, including calculating edit distances for fuzzy matches. We thank the SAINTes team at Microsoft Research for their advisement; Nicole Forsgren and Denae Ford Robinson for advising on questions capturing perceived productivity with the SPACE framework, and Tom Zimmermann and Christian Bird for recommending more time intervals to consider for suggestion fate monitoring. We thank Rahul Pandita for \LaTeX support and proofreading. Finally, we are grateful to GitHub Incorporated for supporting this research.

\onecolumn
\section{Appendix}

\appendix

\section{Summary of usage measurements collected} \label{appendix:usage}

This table shows summary statistics of our core metrics (highlighted in black). We include other possible metrics (arising from different normalization options) in the table for context.

\begin{table*}[h]
\footnotesize
\centering
\begin{tabular}{p{6cm}p{1.5cm}p{1.5cm}p{1.5cm}p{1.5cm}p{1.5cm}p{1.5cm}lrrrrrr}
\toprule
\textbf{Metric}& \textbf{N} & \textbf{Mean} & \textbf{Std.} & \textbf{Min.} & \textbf{Median} & \textbf{Max.}\\
\midrule
\light{{\tt opportunity}} & \light{2,047} & \light{13,085.23} & \light{14,493.16} & \light{1.00} & \light{8,686.00} & \light{241,033.00} \\
\rowcolor{RowGray}{\tt shown } & 2,047 & 1,872.05 & 1,922.22 & 0.00 & 1,276.00 & 15,832.00 \\
\light{{\tt accepted}} & \light{2,047} & \light{503.94} & \light{639.20} & \light{0.00} & \light{293.00} & \light{5,851.00} \\
\rowcolor{RowGray}\light{{\tt unchanged\_30}} & \light{2,047} & \light{328.96} & \light{449.91} & \light{0.00} & \light{178.00} & \light{4,253.00} \\
\light{{\tt unchanged\_120}} & \light{2,047} & \light{289.15} & \light{400.16} & \light{0.00} & \light{155.00} & \light{3,937.00} \\
\rowcolor{RowGray}\light{{\tt unchanged\_300}} & \light{2,047} & \light{262.31} & \light{367.78} & \light{0.00} & \light{140.00} & \light{3,737.00} \\
\light{{\tt unchanged\_600}} & \light{2,047} & \light{240.69} & \light{342.01} & \light{0.00} & \light{125.00} & \light{3,487.00} \\
\rowcolor{RowGray}\light{{\tt accepted\_char}} & \light{2,047} & \light{25,869.83} & \light{33,288.97} & \light{0.00} & \light{14,662.00} & \light{320,080.00} \\
\light{{\tt active\_hour}} & \light{2,047} & \light{77.10} & \light{55.86} & \light{1.00} & \light{66.00} & \light{364.00} \\
\rowcolor{RowGray}\light{{\tt mostly\_unchanged\_30}} & \light{2,047} & \light{434.24} & \light{556.89} & \light{0.00} & \light{254.00} & \light{5,025.00} \\
\light{{\tt mostly\_unchanged\_120}} & \light{2,047} & \light{406.51} & \light{523.54} & \light{0.00} & \light{237.00} & \light{4,785.00} \\
\rowcolor{RowGray}\light{{\tt mostly\_unchanged\_300}} & \light{2,047} & \light{390.58} & \light{503.78} & \light{0.00} & \light{230.00} & \light{4,586.00} \\
\light{{\tt mostly\_unchanged\_600}} & \light{2,047} & \light{382.34} & \light{493.05} & \light{0.00} & \light{222.00} & \light{4,491.00} \\
\rowcolor{RowGray}\light{{\tt opportunity\_per\_active\_hour}} & \light{2,047} & \light{158.07} & \light{107.40} & \light{1.00} & \light{134.40} & \light{1,844.00} \\
{{\tt shown\_per\_active\_hour}} & {2,047} & {22.52} & {13.45} & {0.00} & {20.02} & {137.75} \\
\rowcolor{RowGray}{{\tt accepted\_per\_active\_hour}} & {2,047} & {6.24} & {5.76} & {0.00} & {4.60} & \light{58.56} \\
{\tt shown\_per\_opportunity } & 2,047 & 0.15 & 0.05 & 0.00 & 0.15 & 0.37 \\
\rowcolor{RowGray}{\tt accepted\_per\_opportunity } & 2,047 & 0.04 & 0.03 & 0.00 & 0.04 & 0.22 \\
{\tt accepted\_per\_shown } & 2,038 & 0.26 & 0.12 & 0.00 & 0.24 & 1.00 \\
\rowcolor{RowGray}{{\tt accepted\_char\_per\_active\_hour}} & {2,047} & {335.15} & {464.08} & {0.00} & \light{235.00} & \light{14,064.00} \\
\light{{\tt accepted\_char\_per\_opportunity}} & \light{2,047} & \light{2.14} & \light{1.76} & \light{0.00} & \light{1.71} & \light{20.87} \\
\rowcolor{RowGray}\light{{\tt accepted\_char\_per\_shown}} & \light{2,038} & \light{13.73} & \light{10.00} & \light{0.00} & \light{11.63} & \light{194.00} \\
\light{{\tt accepted\_char\_per\_accepted}} & \light{2,019} & \light{52.84} & \light{22.14} & \light{5.00} & \light{48.85} & \light{494.83} \\
\rowcolor{RowGray}\light{{\tt mostly\_unchanged\_30\_per\_active\_hour}} & \light{2,047} & \light{5.37} & \light{5.02} & \light{0.00} & \light{3.89} & \light{51.92} \\
\light{{\tt mostly\_unchanged\_30\_per\_opportunity}} & \light{2,047} & \light{0.04} & \light{0.02} & \light{0.00} & \light{0.03} & \light{0.19} \\
\rowcolor{RowGray}\light{{\tt mostly\_unchanged\_30\_per\_shown}} & \light{2,038} & \light{0.22} & \light{0.11} & \light{0.00} & \light{0.21} & \light{1.00} \\
{\tt mostly\_unchanged\_30\_per\_accepted } & 2,019 & 0.86 & 0.08 & 0.00 & 0.86 & 1.00 \\
\rowcolor{RowGray}\light{{\tt mostly\_unchanged\_120\_per\_active\_hour}} & \light{2,047} & \light{5.01} & \light{4.66} & \light{0.00} & \light{3.67} & \light{49.00} \\
\light{{\tt mostly\_unchanged\_120\_per\_opportunity}} & \light{2,047} & \light{0.03} & \light{0.02} & \light{0.00} & \light{0.03} & \light{0.19} \\
\rowcolor{RowGray}\light{{\tt mostly\_unchanged\_120\_per\_shown}} & \light{2,038} & \light{0.21} & \light{0.11} & \light{0.00} & \light{0.19} & \light{1.00} \\
{\tt mostly\_unchanged\_120\_per\_accepted } & 2,019 & 0.80 & 0.10 & 0.00 & 0.81 & 1.00 \\
\rowcolor{RowGray}\light{{\tt mostly\_unchanged\_300\_per\_active\_hour}} & \light{2,047} & \light{4.81} & \light{4.40} & \light{0.00} & \light{3.54} & \light{46.61} \\
\light{{\tt mostly\_unchanged\_300\_per\_opportunity}} & \light{2,047} & \light{0.03} & \light{0.02} & \light{0.00} & \light{0.03} & \light{0.19} \\
\rowcolor{RowGray}\light{{\tt mostly\_unchanged\_300\_per\_shown}} & \light{2,038} & \light{0.20} & \light{0.10} & \light{0.00} & \light{0.18} & \light{1.00} \\
{\tt mostly\_unchanged\_300\_per\_accepted } & 2,019 & 0.77 & 0.11 & 0.00 & 0.77 & 1.00 \\
\rowcolor{RowGray}\light{{\tt mostly\_unchanged\_600\_per\_active\_hour}} & \light{2,047} & \light{4.72} & \light{4.32} & \light{0.00} & \light{3.51} & \light{44.85} \\
\light{{\tt mostly\_unchanged\_600\_per\_opportunity}} & \light{2,047} & \light{0.03} & \light{0.02} & \light{0.00} & \light{0.03} & \light{0.19} \\
\rowcolor{RowGray}\light{{\tt mostly\_unchanged\_600\_per\_shown}} & \light{2,038} & \light{0.20} & \light{0.10} & \light{0.00} & \light{0.18} & \light{1.00} \\
{\tt mostly\_unchanged\_600\_per\_accepted } & 2,019 & 0.76 & 0.11 & 0.00 & 0.76 & 1.00 \\
\rowcolor{RowGray}{{\tt unchanged\_30\_per\_active\_hour}} & {2,047} & {4.03} & {4.06} & {0.00} & {2.89} & {43.84} \\
{\tt unchanged\_30\_per\_opportunity } & 2,047 & 0.03 & 0.02 & 0.00 & 0.02 & 0.19 \\
\rowcolor{RowGray}\light{{\tt unchanged\_30\_per\_shown}} & \light{2,038} & \light{0.17} & \light{0.10} & \light{0.00} & \light{0.15} & \light{0.64} \\
{\tt unchanged\_30\_per\_accepted } & 2,019 & 0.64 & 0.17 & 0.00 & 0.67 & 1.00 \\
\rowcolor{RowGray}{{\tt unchanged\_120\_per\_active\_hour}} & {2,047} & {3.51} & {3.50} & {0.00} & {2.52} & {40.00} \\
{\tt unchanged\_120\_per\_opportunity } & 2,047 & 0.02 & 0.02 & 0.00 & 0.02 & 0.19 \\
\rowcolor{RowGray}\light{{\tt unchanged\_120\_per\_shown}} & \light{2,038} & \light{0.15} & \light{0.09} & \light{0.00} & \light{0.13} & \light{0.63} \\
{\tt unchanged\_120\_per\_accepted } & 2,019 & 0.56 & 0.16 & 0.00 & 0.59 & 1.00 \\
\rowcolor{RowGray}{{\tt unchanged\_300\_per\_active\_hour}} & {2,047} & {3.13} & {3.08} & {0.00} & {2.24} & {36.77} \\
{\tt unchanged\_300\_per\_opportunity } & 2,047 & 0.02 & 0.02 & 0.00 & 0.02 & 0.17 \\
\rowcolor{RowGray}\light{{\tt unchanged\_300\_per\_shown}} & \light{2,038} & \light{0.13} & \light{0.08} & \light{0.00} & \light{0.12} & \light{0.62} \\
{\tt unchanged\_300\_per\_accepted } & 2,019 & 0.51 & 0.16 & 0.00 & 0.53 & 1.00 \\
\rowcolor{RowGray}{{\tt unchanged\_600\_per\_active\_hour}} & {2,047} & {2.82} & {2.76} & {0.00} & {2.06} & {34.15} \\
{\tt unchanged\_600\_per\_opportunity } & 2,047 & 0.02 & 0.02 & 0.00 & 0.02 & 0.17 \\
\rowcolor{RowGray}\light{{\tt unchanged\_600\_per\_shown}} & \light{2,038} & \light{0.12} & \light{0.07} & \light{0.00} & \light{0.11} & \light{0.61} \\
{\tt unchanged\_600\_per\_accepted } & 2,019 & 0.46 & 0.16 & 0.00 & 0.48 & 1.00 \\
\bottomrule
\end{tabular}
\end{table*}

%\afterpage{\clearpage}

\section{Correlations between usage measurements and survey questions}
\label{appendix:correlation}

This table shows the correlation of the aggregate productivity score from our survey against our core metrics (highlighted in black), as well as their PLS scores. We include other possible metrics (arising from different normalization options) in the table for context.

\begin{table*}[h]
\footnotesize

    \begin{tabular}{p{8cm}p{1.5cm}p{1.5cm}p{1.5cm}p{1.5cm}p{1.5cm}lrrrrr}
\toprule
\textbf{Metric}& \textbf{N} & \textbf{Coefficient} & \textbf{P-Value} & \textbf{PLS-1} & \textbf{PLS-2} \\
\midrule
{\tt accepted\_per\_shown} & 1,780 & 0.24 & <0.0001 & 0.0133 & 0.029\\
\rowcolor{RowGray}\light{{\tt mostly\_unchanged\_30\_per\_shown}} & \light{1,780} & \light{0.23} & \light{<0.0001} & &\\
\light{{\tt mostly\_unchanged\_120\_per\_shown}} & \light{1,780} & \light{0.23} & \light{<0.0001} \\
\rowcolor{RowGray}\light{{\tt mostly\_unchanged\_300\_per\_shown}} & \light{1,780} & \light{0.22} & \light{<0.0001} & &\\
{\tt accepted\_per\_opportunity} & 1,789 & 0.22 & <0.0001 & 0.0122 & 0.0204\\
\rowcolor{RowGray}\light{{\tt mostly\_unchanged\_600\_per\_shown}} & \light{1,780} & \light{0.22} & \light{<0.0001} & &\\
\light{{\tt unchanged\_30\_per\_shown}} & \light{1,780} & \light{0.21} & \light{<0.0001} \\
\rowcolor{RowGray}\light{{\tt mostly\_unchanged\_30\_per\_opportunity}} & \light{1,789} & \light{0.21} & \light{<0.0001} & &\\
\light{{\tt mostly\_unchanged\_120\_per\_opportunity}} & \light{1,789} & \light{0.21} & \light{<0.0001} \\
\rowcolor{RowGray}\light{{\tt mostly\_unchanged\_30\_per\_active\_hour}} & \light{1,789} & \light{0.21} & \light{<0.0001} & &\\
{\tt accepted\_per\_active\_hour} & 1,789 & 0.21 & <0.0001 & 0.0116 & 0.0164\\
\rowcolor{RowGray}\light{{\tt mostly\_unchanged\_120\_per\_active\_hour}} & \light{1,789} & \light{0.21} & \light{<0.0001} & &\\
\light{{\tt unchanged\_120\_per\_shown}} & \light{1,780} & \light{0.21} & \light{<0.0001} \\
\rowcolor{RowGray}\light{{\tt mostly\_unchanged\_300\_per\_active\_hour}} & \light{1,789} & \light{0.21} & \light{<0.0001} & &\\
\light{{\tt mostly\_unchanged\_300\_per\_opportunity}} & \light{1,789} & \light{0.21} & \light{<0.0001} \\
\rowcolor{RowGray}\light{{\tt mostly\_unchanged\_600\_per\_active\_hour}} & \light{1,789} & \light{0.20} & \light{<0.0001} & &\\
\light{{\tt mostly\_unchanged\_600\_per\_opportunity}} & \light{1,789} & \light{0.20} & \light{<0.0001} \\
\rowcolor{RowGray}{\tt unchanged\_30\_per\_opportunity} & 1,789 & 0.20 & <0.0001 & 0.0112 & 0.0122\\
{{\tt unchanged\_30\_per\_active\_hour}} & {1,789} & {0.20} & {<0.0001} & 0.0111 & 0.0104\\
\rowcolor{RowGray}\light{{\tt unchanged\_300\_per\_shown}} & \light{1,780} & \light{0.19} & \light{<0.0001} & &\\
{\tt unchanged\_120\_per\_active\_hour} & {1,789} & {0.19} & {<0.0001} & 0.011 & 0.0092\\
\rowcolor{RowGray}{\tt unchanged\_120\_per\_opportunity} & 1,789 & 0.19 & <0.0001 & 0.0108 & 0.0099\\
\light{{\tt accepted\_char\_per\_opportunity}} & \light{1,789} & \light{0.19} & \light{<0.0001} & &\\
\rowcolor{RowGray}{{\tt unchanged\_300\_per\_active\_hour}} & {1,789} & {0.19} & {<0.0001} & 0.0107 & 0.0082\\
{\tt unchanged\_300\_per\_opportunity} & 1,789 & 0.18 & <0.0001 & 0.0103 & 0.0082\\
\rowcolor{RowGray}{{\tt unchanged\_600\_per\_active\_hour}} & {1,789} & {0.18} & {<0.0001} & 0.01 & 0.006\\
\light{{\tt accepted\_char\_per\_shown}} & \light{1,780} & \light{0.17} & \light{<0.0001} & &\\
\rowcolor{RowGray}\light{{\tt unchanged\_600\_per\_shown}} & \light{1,780} & \light{0.17} & \light{<0.0001} & &\\
{\tt unchanged\_600\_per\_opportunity} & 1,789 & 0.16 & <0.0001 & 0.0093 & 0.0049\\
\rowcolor{RowGray}{{\tt accepted\_char\_per\_active\_hour}} & {1,789} & {0.16} & {<0.0001} & 0.0094 & 0.0155\\
\light{{\tt accepted\_char}} & \light{1,789} & \light{0.11} & \light{<0.0001} & &\\
\rowcolor{RowGray}\light{{\tt mostly\_unchanged\_30}} & \light{1,789} & \light{0.11} & \light{<0.0001} & &\\
{{\tt shown\_per\_active\_hour}} & {1,789} & {0.11} & {<0.0001} & 0.006 & 0.0025\\
\rowcolor{RowGray}\light{{\tt accepted}} & \light{1,789} & \light{0.11} & \light{<0.0001} & &\\
\light{{\tt mostly\_unchanged\_120}} & \light{1,789} & \light{0.11} & \light{<0.0001} & &\\
\rowcolor{RowGray}\light{{\tt mostly\_unchanged\_600}} & \light{1,789} & \light{0.11} & \light{<0.0001} & &\\
{\tt shown\_per\_opportunity} & 1,789 & 0.11 & <0.0001 & 0.0059 & 0.0082\\
\rowcolor{RowGray}\light{{\tt mostly\_unchanged\_300}} & \light{1,789} & \light{0.11} & \light{<0.0001} & &\\
\light{{\tt unchanged\_30}} & \light{1,789} & \light{0.11} & \light{<0.0001} & &\\
\rowcolor{RowGray}\light{{\tt unchanged\_120}} & \light{1,789} & \light{0.10} & \light{<0.0001} & &\\
\light{{\tt unchanged\_300}} & \light{1,789} & \light{0.10} & \light{<0.0001} & &\\
\rowcolor{RowGray}\light{{\tt unchanged\_600}} & \light{1,789} & \light{0.10} & \light{<0.0001} & &\\
{\tt mostly\_unchanged\_30\_per\_accepted} & 1,763 & 0.07 & 0.01 & 0.0036 & 0.0077\\
\rowcolor{RowGray}{\tt unchanged\_30\_per\_accepted} & 1,763 & 0.06 & 0.02 & 0.0031 & -0.0011\\
{\tt mostly\_unchanged\_120\_per\_accepted} & 1,763 & 0.05 & 0.04 & 0.0028 & 0.0064\\
\rowcolor{RowGray}{\tt unchanged\_120\_per\_accepted} & 1,763 & 0.04 & 0.09 & 0.0023 & -0.0047\\
{\tt mostly\_unchanged\_600\_per\_accepted} & 1,763 & 0.04 & 0.09 & 0.0022 & 0.007\\
\rowcolor{RowGray}{\tt mostly\_unchanged\_300\_per\_accepted} & 1,763 & 0.04 & 0.09 & 0.0022 & 0.0053\\
\light{{\tt accepted\_char\_per\_accepted}} & \light{1,763} & \light{0.03} & \light{0.20} \\
\rowcolor{RowGray}\light{{\tt opportunity\_per\_active\_hour}} & \light{1,789} & \light{0.03} & \light{0.24} & &\\
{\tt unchanged\_300\_per\_accepted} & 1,763 & 0.02 & 0.40 & 0.0011 & -0.0094\\
\rowcolor{RowGray}{\tt shown} & 1,789 & 0.01 & 0.75 & 4e-04 & -0.0149\\
{\tt unchanged\_600\_per\_accepted} & 1,763 & -0.00 & 0.95 & -1e-04 & -0.0138\\
\rowcolor{RowGray}\light{{\tt opportunity}} & \light{1,789} & \light{-0.04} & \light{0.08} & &\\
\light{{\tt active\_hour}} & \light{1,789} & \light{-0.05} & \light{0.03} \\
\bottomrule
    \end{tabular}
\end{table*}

%\afterpage{\clearpage}

\section{Aspects of Productivity Measured In The Survey}
\label{appendix:surveyaspects}

This table shows the relationship between the survey statements, the metrics and the different dimension of the SPACE framework~\cite{forsgren2021space}.

\begingroup
\setlength{\tabcolsep}{10pt} % Default value: 6pt
\renewcommand{\arraystretch}{1.7} % Default value: 1

\begin{table*}[h]
\footnotesize
    \centering
\begin{tabular}{p{8cm}p{3cm}p{1cm}p{3cm}llcl}
\toprule
\textbf{Survey statements}                                                                                 & \textbf{Productivity aspect}                                                  & \textbf{Code}       & \textbf{Metric name}                            \\
\midrule
``I am more productive when using \Copilot''                                                            & Perceived productivity                                                        &                     & \texttt{more\_productive}         \\
\hline
``I feel more fulfilled with my job when using \Copilot.''                                              &                                                                               &                     & \texttt{more\_fulfilled}          \\
``I find myself less frustrated during coding sessions when using \Copilot.''                           &                                                                               &                     & \texttt{less\_frustrated}         \\
``I can focus on more satisfying work when using \Copilot.''                                            & \multirow{-3}{*}{\parbox[t]{3cm}{Satisfaction and well-being}}                         & \multirow{-3}{*}{S} & \texttt{focus\_satisfying}        \\
\hline
``While working with an unfamiliar language, I make progress faster when using \Copilot.''              &                                                                               &                     & \texttt{unfamiliar\_progress}     \\
``The code I write using \Copilot is better than the code I would have written without \Copilot.'' & \multirow{-2}{*}{Performance}                                         & \multirow{-2}{*}{P} & \texttt{better\_code}             \\
\hline
n/a                                                                                                        & Activity                                                              & A                   & n/a                                             \\
\hline
``I learn from the suggestions \Copilot shows me.''                                                     & Communication and collaboration~\cite{wang2020human} & C                   & \texttt{learn\_from}              \\
\hline
``Using \Copilot helps me stay in the flow.''                                                           &                                                                               &                     & \texttt{stay\_in\_flow}           \\
``I complete tasks faster when using \Copilot.''                                                        &                                                                               &                     & \texttt{tasks\_faster}            \\
``I complete repetitive programming tasks faster when using \Copilot.''                                 &                                                                               &                     & \texttt{repetitive\_faster}       \\
``I spend less mental effort on repetitive programming tasks when using \Copilot.''                     &                                                                               &                     & \texttt{less\_effort\_repetitive} \\
``I spend less time searching for information or examples when using \Copilot.''                        & \multirow{-5}{*}{\parbox[t]{3cm}{Efficiency and flow}}                                 & \multirow{-5}{*}{E} & \texttt{less\_time\_searching}   \\
\bottomrule
\end{tabular}
\end{table*}

\endgroup
\clearpage
\section{Incremental Feature Selection}

Univariate fit to \texttt{aggregate\_productivity}, excluding 120, 300 and 600 second unchanged metrics:

\begin{table*}[h]
\small
    \centering
\begin{tabular}{lrrr}
\toprule
\textbf{Behavioral Metric}           & \multicolumn{1}{l}{\textbf{Coefficient}} & \multicolumn{1}{l}{\textbf{P-value}} & \multicolumn{1}{l}{\textbf{SSR}} \\
\hline
\texttt{accepted\_per\_shown}                 & 0.13                                     & 0.00                                 & 466.26                           \\
\rowcolor{RowGray}\texttt{accepted\_per\_opportunity}           & 0.12                                     & 0.00                                 & 471.28                           \\
\texttt{accepted\_per\_hour  }                & 0.11                                     & 0.00                                 & 473.79                           \\
\rowcolor{RowGray}\texttt{unchanged\_30\_per\_opportunity}      & 0.11                                     & 0.00                                 & 475.43                           \\
\texttt{unchanged\_30\_per\_hour}             & 0.10                                     & 0.00                                 & 476.36                           \\
\rowcolor{RowGray}\texttt{accepted\_char\_per\_hour}            & 0.09                                     & 0.00                                 & 482.39                           \\
\texttt{shown\_per\_opportunity}              & 0.06                                     & 0.00                                 & 489.60                           \\
\rowcolor{RowGray}\texttt{shown\_per\_hour}                     & 0.06                                     & 0.00                                 & 489.66                           \\
\texttt{mostly\_unchanged\_30\_per\_accepted} & 0.03                                     & 0.01                                 & 493.49                           \\
\rowcolor{RowGray}\texttt{unchanged\_30\_per\_accepted}         & 0.03                                     & 0.02                                 & 494.04                           \\
\texttt{shown}                                & 0.00                                     & 0.78                                 & 495.58                           \\
\bottomrule
\end{tabular}
\end{table*}

Best metric fit to a univariate model modeling residuals of \texttt{accepted\_per\_shown} univariate model:

\begin{table*}[h]
\small
    \centering
\begin{tabular}{lrrr}
\toprule
\textbf{Behavioral Metric}           & \multicolumn{1}{l}{\textbf{Coefficient}} & \multicolumn{1}{l}{\textbf{P-value}} & \multicolumn{1}{l}{\textbf{SSR}} \\
\hline
\texttt{shown\_per\_hour}                     & 0.03                                     & 0.0363                               & 465.10                           \\
\rowcolor{RowGray}\texttt{accepted\_char\_per\_hour}            & 0.03                                     & 0.0385                               & 465.13                           \\
\texttt{shown\_per\_opportunity}              & 0.03                                     & 0.0399                               & 465.15                           \\
\rowcolor{RowGray}\texttt{accepted\_per\_hour}                  & 0.02                                     & 0.0465                               & 465.21                           \\
\texttt{unchanged\_30\_per\_hour}             & 0.02                                     & 0.0955                               & 465.53                           \\
\rowcolor{RowGray}\texttt{mostly\_unchanged\_30\_per\_accepted} & 0.02                                     & 0.2124                               & 465.85                           \\
\texttt{unchanged\_30\_per\_accepted}         & 0.01                                     & 0.2461                               & 465.91                           \\
\rowcolor{RowGray}\texttt{accepted\_per\_opportunity}           & 0.01                                     & 0.4819                               & 466.13                           \\
\texttt{unchanged\_30\_per\_opportunity}      & 0.01                                     & 0.5442                               & 466.17                           \\
\rowcolor{RowGray}\texttt{shown}                                & -0.01                                    & 0.5539                               & 466.17                           \\
\bottomrule
\end{tabular}
\end{table*}
\clearpage
\section{Publicly posted comments}
\label{appendix:comments}
Below we include a selection of (unsolicited) publicly posted comments which give a sense that developers are thinking about \emph{engagement} with \Copilot rather than solely the immediate correctness of a completion.

\begin{quote}
    It's a little like pair programming with an incredibly eager junior developer who has read a lot of the documentation of every popular API in the world. 
    \flushright \url{https://news.ycombinator.com/item?id=30691608}
\end{quote}
\vspace{1cm}
\begin{quote}
    Cycling through \Copilot suggestions and manually editing the suggested code is an amazing flow. What I really like is that \GitHubCopilot adapts to my own code style.
    \flushright \url{https://www.meetup.com/Microsoft-Reactor-Toronto/events/284609940/}
\end{quote}
\vspace{1cm}
\begin{quote}
    It says, `How can I facilitate your thinking process?' rather than, `How can I take away your thinking process and just give you code?'
    \flushright \url{https://www.protocol.com/workplace/github-copilot-ai-developers}
\end{quote}
\vspace{1cm}
\begin{quote}
    i was writing a function that evaluates a polynomial, in a lambda (`let eval\_polynomial = |`) and it autofilled a function that evaluated the polynomial but i wanted horner's method, so i deleted and typed `let eval\_polynomial\_horner = |`it correctly autofilled (with one small error) horner's method for evaluating polynomials
    \flushright \url{https://twitter.com/dystopiabreaker/status/1488692230114070529}
\end{quote}
\vspace{1cm}
\begin{quote}
    Just pasted in an AttributeError as a comment in my Python file, and \Copilot began trying to help me debug my implementation of a  @HuggingFace  Transformers model. Its advice wasn't 100\% all-knowing, but was enough to get me to a resolution!
    \flushright \url{https://twitter.com/DynamicWebPaige/status/1447587091995500546}
\end{quote}

%\afterpage{\clearpage}
\bibliographystyle{ACM-Reference-Format}
\bibliography{references}


\begin{thebibliography}{21}

%%% the Web address.

\ifx \showCODEN    \undefined \def \showCODEN     #1{\unskip}     \fi
\ifx \showDOI      \undefined \def \showDOI       #1{#1}\fi
\ifx \showISBNx    \undefined \def \showISBNx     #1{\unskip}     \fi
\ifx \showISBNxiii \undefined \def \showISBNxiii  #1{\unskip}     \fi
\ifx \showISSN     \undefined \def \showISSN      #1{\unskip}     \fi
\ifx \showLCCN     \undefined \def \showLCCN      #1{\unskip}     \fi
\ifx \shownote     \undefined \def \shownote      #1{#1}          \fi
\ifx \showarticletitle \undefined \def \showarticletitle #1{#1}   \fi
\ifx \showURL      \undefined \def \showURL       {\relax}        \fi
% invisible to TeX
\providecommand\bibfield[2]{#2}
\providecommand\bibinfo[2]{#2}
\providecommand\natexlab[1]{#1}
\providecommand\showeprint[2][]{arXiv:#2}

\bibitem[Amann et~al\mbox{.}(2016)]%
        {DBLP:conf/wcre/AmannPNM16}
\bibfield{author}{\bibinfo{person}{Sven Amann}, \bibinfo{person}{Sebastian
  Proksch}, \bibinfo{person}{Sarah Nadi}, {and} \bibinfo{person}{Mira Mezini}.}
  \bibinfo{year}{2016}\natexlab{}.
\newblock \showarticletitle{A Study of Visual Studio Usage in Practice}. In
  \bibinfo{booktitle}{\emph{{IEEE} 23rd International Conference on Software
  Analysis, Evolution, and Reengineering, {SANER} 2016, Suita, Osaka, Japan,
  March 14-18, 2016 - Volume 1}}. \bibinfo{publisher}{{IEEE} Computer Society},
  \bibinfo{pages}{124--134}.
\newblock
\urldef\tempurl%
\url{https://doi.org/10.1109/SANER.2016.39}
\showDOI{\tempurl}


\bibitem[Austin et~al\mbox{.}(2021)]%
        {DBLP:journals/corr/abs-2108-07732}
\bibfield{author}{\bibinfo{person}{Jacob Austin}, \bibinfo{person}{Augustus
  Odena}, \bibinfo{person}{Maxwell Nye}, \bibinfo{person}{Maarten Bosma},
  \bibinfo{person}{Henryk Michalewski}, \bibinfo{person}{David Dohan},
  \bibinfo{person}{Ellen Jiang}, \bibinfo{person}{Carrie~J. Cai},
  \bibinfo{person}{Michael Terry}, \bibinfo{person}{Quoc~V. Le}, {and}
  \bibinfo{person}{Charles Sutton}.} \bibinfo{year}{2021}\natexlab{}.
\newblock \showarticletitle{Program Synthesis with Large Language Models}.
\newblock \bibinfo{journal}{\emph{CoRR}}  \bibinfo{volume}{abs/2108.07732}
  (\bibinfo{year}{2021}).
\newblock
\showeprint[arXiv]{2108.07732}
\urldef\tempurl%
\url{https://arxiv.org/abs/2108.07732}
\showURL{%
\tempurl}


\bibitem[Aye et~al\mbox{.}(2021)]%
        {DBLP:conf/icse/AyeKL21}
\bibfield{author}{\bibinfo{person}{Gareth~Ari Aye}, \bibinfo{person}{Seohyun
  Kim}, {and} \bibinfo{person}{Hongyu Li}.} \bibinfo{year}{2021}\natexlab{}.
\newblock \showarticletitle{Learning Autocompletion from Real-World Datasets}.
  In \bibinfo{booktitle}{\emph{43rd {IEEE/ACM} International Conference on
  Software Engineering: Software Engineering in Practice, {ICSE} {(SEIP)} 2021,
  Madrid, Spain, May 25-28, 2021}}. \bibinfo{publisher}{{IEEE}},
  \bibinfo{pages}{131--139}.
\newblock
\urldef\tempurl%
\url{https://doi.org/10.1109/ICSE-SEIP52600.2021.00022}
\showDOI{\tempurl}


\bibitem[Beller et~al\mbox{.}(2020)]%
        {beller2020mind}
\bibfield{author}{\bibinfo{person}{Moritz Beller}, \bibinfo{person}{Vince
  Orgovan}, \bibinfo{person}{Spencer Buja}, {and} \bibinfo{person}{Thomas
  Zimmermann}.} \bibinfo{year}{2020}\natexlab{}.
\newblock \showarticletitle{Mind the gap: on the relationship between
  automatically measured and self-reported productivity}.
\newblock \bibinfo{journal}{\emph{IEEE Software}} \bibinfo{volume}{38},
  \bibinfo{number}{5} (\bibinfo{year}{2020}), \bibinfo{pages}{24--31}.
\newblock


\bibitem[Chen et~al\mbox{.}(2021)]%
        {DBLP:journals/corr/abs-2107-03374}
\bibfield{author}{\bibinfo{person}{Mark Chen}, \bibinfo{person}{Jerry Tworek},
  \bibinfo{person}{Heewoo Jun}, \bibinfo{person}{Qiming Yuan},
  \bibinfo{person}{Henrique~Ponde de Oliveira~Pinto}, \bibinfo{person}{Jared
  Kaplan}, \bibinfo{person}{Harrison Edwards}, \bibinfo{person}{Yuri Burda},
  \bibinfo{person}{Nicholas Joseph}, \bibinfo{person}{Greg Brockman},
  \bibinfo{person}{Alex Ray}, \bibinfo{person}{Raul Puri},
  \bibinfo{person}{Gretchen Krueger}, \bibinfo{person}{Michael Petrov},
  \bibinfo{person}{Heidy Khlaaf}, \bibinfo{person}{Girish Sastry},
  \bibinfo{person}{Pamela Mishkin}, \bibinfo{person}{Brooke Chan},
  \bibinfo{person}{Scott Gray}, \bibinfo{person}{Nick Ryder},
  \bibinfo{person}{Mikhail Pavlov}, \bibinfo{person}{Alethea Power},
  \bibinfo{person}{Lukasz Kaiser}, \bibinfo{person}{Mohammad Bavarian},
  \bibinfo{person}{Clemens Winter}, \bibinfo{person}{Philippe Tillet},
  \bibinfo{person}{Felipe~Petroski Such}, \bibinfo{person}{Dave Cummings},
  \bibinfo{person}{Matthias Plappert}, \bibinfo{person}{Fotios Chantzis},
  \bibinfo{person}{Elizabeth Barnes}, \bibinfo{person}{Ariel Herbert{-}Voss},
  \bibinfo{person}{William~Hebgen Guss}, \bibinfo{person}{Alex Nichol},
  \bibinfo{person}{Alex Paino}, \bibinfo{person}{Nikolas Tezak},
  \bibinfo{person}{Jie Tang}, \bibinfo{person}{Igor Babuschkin},
  \bibinfo{person}{Suchir Balaji}, \bibinfo{person}{Shantanu Jain},
  \bibinfo{person}{William Saunders}, \bibinfo{person}{Christopher Hesse},
  \bibinfo{person}{Andrew~N. Carr}, \bibinfo{person}{Jan Leike},
  \bibinfo{person}{Joshua Achiam}, \bibinfo{person}{Vedant Misra},
  \bibinfo{person}{Evan Morikawa}, \bibinfo{person}{Alec Radford},
  \bibinfo{person}{Matthew Knight}, \bibinfo{person}{Miles Brundage},
  \bibinfo{person}{Mira Murati}, \bibinfo{person}{Katie Mayer},
  \bibinfo{person}{Peter Welinder}, \bibinfo{person}{Bob McGrew},
  \bibinfo{person}{Dario Amodei}, \bibinfo{person}{Sam McCandlish},
  \bibinfo{person}{Ilya Sutskever}, {and} \bibinfo{person}{Wojciech Zaremba}.}
  \bibinfo{year}{2021}\natexlab{}.
\newblock \showarticletitle{Evaluating Large Language Models Trained on Code}.
\newblock \bibinfo{journal}{\emph{CoRR}}  \bibinfo{volume}{abs/2107.03374}
  (\bibinfo{year}{2021}).
\newblock
\showeprint[arXiv]{2107.03374}
\urldef\tempurl%
\url{https://arxiv.org/abs/2107.03374}
\showURL{%
\tempurl}


\bibitem[Forsgren et~al\mbox{.}(2021)]%
        {forsgren2021space}
\bibfield{author}{\bibinfo{person}{Nicole Forsgren},
  \bibinfo{person}{Margaret-Anne Storey}, \bibinfo{person}{Chandra Maddila},
  \bibinfo{person}{Thomas Zimmermann}, \bibinfo{person}{Brian Houck}, {and}
  \bibinfo{person}{Jenna Butler}.} \bibinfo{year}{2021}\natexlab{}.
\newblock \showarticletitle{The SPACE of Developer Productivity: There's more
  to it than you think.}
\newblock \bibinfo{journal}{\emph{Queue}} \bibinfo{volume}{19},
  \bibinfo{number}{1} (\bibinfo{year}{2021}), \bibinfo{pages}{20--48}.
\newblock


\bibitem[Hellendoorn et~al\mbox{.}(2019)]%
        {DBLP:conf/icse/HellendoornPGB19}
\bibfield{author}{\bibinfo{person}{Vincent~J. Hellendoorn},
  \bibinfo{person}{Sebastian Proksch}, \bibinfo{person}{Harald~C. Gall}, {and}
  \bibinfo{person}{Alberto Bacchelli}.} \bibinfo{year}{2019}\natexlab{}.
\newblock \showarticletitle{When code completion fails: a case study on
  real-world completions}. In \bibinfo{booktitle}{\emph{Proceedings of the 41st
  International Conference on Software Engineering, {ICSE} 2019, Montreal, QC,
  Canada, May 25-31, 2019}}, \bibfield{editor}{\bibinfo{person}{Joanne~M.
  Atlee}, \bibinfo{person}{Tevfik Bultan}, {and} \bibinfo{person}{Jon Whittle}}
  (Eds.). \bibinfo{publisher}{{IEEE} / {ACM}}, \bibinfo{pages}{960--970}.
\newblock
\urldef\tempurl%
\url{https://doi.org/10.1109/ICSE.2019.00101}
\showDOI{\tempurl}


\bibitem[Hendrycks et~al\mbox{.}(2021)]%
        {DBLP:journals/corr/abs-2105-09938}
\bibfield{author}{\bibinfo{person}{Dan Hendrycks}, \bibinfo{person}{Steven
  Basart}, \bibinfo{person}{Saurav Kadavath}, \bibinfo{person}{Mantas Mazeika},
  \bibinfo{person}{Akul Arora}, \bibinfo{person}{Ethan Guo},
  \bibinfo{person}{Collin Burns}, \bibinfo{person}{Samir Puranik},
  \bibinfo{person}{Horace He}, \bibinfo{person}{Dawn Song}, {and}
  \bibinfo{person}{Jacob Steinhardt}.} \bibinfo{year}{2021}\natexlab{}.
\newblock \showarticletitle{Measuring Coding Challenge Competence With {APPS}}.
\newblock \bibinfo{journal}{\emph{CoRR}}  \bibinfo{volume}{abs/2105.09938}
  (\bibinfo{year}{2021}).
\newblock
\showeprint[arXiv]{2105.09938}
\urldef\tempurl%
\url{https://arxiv.org/abs/2105.09938}
\showURL{%
\tempurl}


\bibitem[Kulal et~al\mbox{.}(2019)]%
        {DBLP:conf/nips/KulalPC0PAL19}
\bibfield{author}{\bibinfo{person}{Sumith Kulal}, \bibinfo{person}{Panupong
  Pasupat}, \bibinfo{person}{Kartik Chandra}, \bibinfo{person}{Mina Lee},
  \bibinfo{person}{Oded Padon}, \bibinfo{person}{Alex Aiken}, {and}
  \bibinfo{person}{Percy Liang}.} \bibinfo{year}{2019}\natexlab{}.
\newblock \showarticletitle{SPoC: Search-based Pseudocode to Code}. In
  \bibinfo{booktitle}{\emph{Advances in Neural Information Processing Systems
  32: Annual Conference on Neural Information Processing Systems 2019, NeurIPS
  2019, December 8-14, 2019, Vancouver, BC, Canada}},
  \bibfield{editor}{\bibinfo{person}{Hanna~M. Wallach}, \bibinfo{person}{Hugo
  Larochelle}, \bibinfo{person}{Alina Beygelzimer}, \bibinfo{person}{Florence
  d'Alch{\'{e}}{-}Buc}, \bibinfo{person}{Emily~B. Fox}, {and}
  \bibinfo{person}{Roman Garnett}} (Eds.). \bibinfo{pages}{11883--11894}.
\newblock
\urldef\tempurl%
\url{https://proceedings.neurips.cc/paper/2019/hash/7298332f04ac004a0ca44cc69ecf6f6b-Abstract.html}
\showURL{%
\tempurl}


\bibitem[Li et~al\mbox{.}(2022)]%
        {alphacode}
\bibfield{author}{\bibinfo{person}{Yujia Li}, \bibinfo{person}{David Choi},
  \bibinfo{person}{Junyoung Chung}, \bibinfo{person}{Nate Kushman},
  \bibinfo{person}{Julian Schrittwieser}, \bibinfo{person}{Rémi Leblond},
  \bibinfo{person}{Tom Eccles}, \bibinfo{person}{James Keeling},
  \bibinfo{person}{Felix Gimeno}, \bibinfo{person}{Agustin Dal~Lago},
  \bibinfo{person}{Thomas Hubert}, \bibinfo{person}{Peter Choy},
  \bibinfo{person}{Cyprien de Masson~d'Autume}, \bibinfo{person}{Igor
  Babuschkin}, \bibinfo{person}{Xinyun Chen}, \bibinfo{person}{Po-Sen Huang},
  \bibinfo{person}{Johannes Welbl}, \bibinfo{person}{Sven Gowal},
  \bibinfo{person}{Alexey Cherepanov}, \bibinfo{person}{James Molloy},
  \bibinfo{person}{Daniel Mankowitz}, \bibinfo{person}{Esme Sutherland~Robson},
  \bibinfo{person}{Pushmeet Kohli}, \bibinfo{person}{Nando de Freitas},
  \bibinfo{person}{Koray Kavukcuoglu}, {and} \bibinfo{person}{Oriol Vinyals}.}
  \bibinfo{year}{2022}\natexlab{}.
\newblock \bibinfo{title}{Competition-Level Code Generation with AlphaCode}.
\newblock
\newblock


\bibitem[See et~al\mbox{.}(2019)]%
        {DBLP:conf/naacl/SeeRKW19}
\bibfield{author}{\bibinfo{person}{Abigail See}, \bibinfo{person}{Stephen
  Roller}, \bibinfo{person}{Douwe Kiela}, {and} \bibinfo{person}{Jason
  Weston}.} \bibinfo{year}{2019}\natexlab{}.
\newblock \showarticletitle{What makes a good conversation? How controllable
  attributes affect human judgments}. In \bibinfo{booktitle}{\emph{Proceedings
  of the 2019 Conference of the North American Chapter of the Association for
  Computational Linguistics: Human Language Technologies, {NAACL-HLT} 2019,
  Minneapolis, MN, USA, June 2-7, 2019, Volume 1 (Long and Short Papers)}},
  \bibfield{editor}{\bibinfo{person}{Jill Burstein}, \bibinfo{person}{Christy
  Doran}, {and} \bibinfo{person}{Thamar Solorio}} (Eds.).
  \bibinfo{publisher}{Association for Computational Linguistics},
  \bibinfo{pages}{1702--1723}.
\newblock
\urldef\tempurl%
\url{https://doi.org/10.18653/v1/n19-1170}
\showDOI{\tempurl}


\bibitem[Svyatkovskiy et~al\mbox{.}(2020)]%
        {DBLP:conf/sigsoft/SvyatkovskiyDFS20}
\bibfield{author}{\bibinfo{person}{Alexey Svyatkovskiy},
  \bibinfo{person}{Shao~Kun Deng}, \bibinfo{person}{Shengyu Fu}, {and}
  \bibinfo{person}{Neel Sundaresan}.} \bibinfo{year}{2020}\natexlab{}.
\newblock \showarticletitle{IntelliCode compose: code generation using
  transformer}. In \bibinfo{booktitle}{\emph{{ESEC/FSE} '20: 28th {ACM} Joint
  European Software Engineering Conference and Symposium on the Foundations of
  Software Engineering, Virtual Event, USA, November 8-13, 2020}},
  \bibfield{editor}{\bibinfo{person}{Prem Devanbu}, \bibinfo{person}{Myra~B.
  Cohen}, {and} \bibinfo{person}{Thomas Zimmermann}} (Eds.).
  \bibinfo{publisher}{{ACM}}, \bibinfo{pages}{1433--1443}.
\newblock
\urldef\tempurl%
\url{https://doi.org/10.1145/3368089.3417058}
\showDOI{\tempurl}


\bibitem[Svyatkovskiy et~al\mbox{.}(2021)]%
        {DBLP:conf/msr/SvyatkovskiyLHR21}
\bibfield{author}{\bibinfo{person}{Alexey Svyatkovskiy},
  \bibinfo{person}{Sebastian Lee}, \bibinfo{person}{Anna Hadjitofi},
  \bibinfo{person}{Maik Riechert}, \bibinfo{person}{Juliana~Vicente Franco},
  {and} \bibinfo{person}{Miltiadis Allamanis}.}
  \bibinfo{year}{2021}\natexlab{}.
\newblock \showarticletitle{Fast and Memory-Efficient Neural Code Completion}.
  In \bibinfo{booktitle}{\emph{18th {IEEE/ACM} International Conference on
  Mining Software Repositories, {MSR} 2021, Madrid, Spain, May 17-19, 2021}}.
  \bibinfo{publisher}{{IEEE}}, \bibinfo{pages}{329--340}.
\newblock
\urldef\tempurl%
\url{https://doi.org/10.1109/MSR52588.2021.00045}
\showDOI{\tempurl}


\bibitem[Vaithilingam et~al\mbox{.}(2022a)]%
        {evalcopilot}
\bibfield{author}{\bibinfo{person}{Priyan Vaithilingam},
  \bibinfo{person}{Tianyi Zhang}, {and} \bibinfo{person}{Elena Glassman}.}
  \bibinfo{year}{2022}\natexlab{a}.
\newblock \showarticletitle{Expectation vs. Experience: Evaluating the
  Usability of Code Generation Tools Powered by Large Language Models}. In
  \bibinfo{booktitle}{\emph{CHI '22 Late-Breaking Work: Proceedings of the 2022
  Conference on Human Factors in Computing Systems}}.
\newblock


\bibitem[Vaithilingam et~al\mbox{.}(2022b)]%
        {10.1145/3491101.3519665}
\bibfield{author}{\bibinfo{person}{Priyan Vaithilingam},
  \bibinfo{person}{Tianyi Zhang}, {and} \bibinfo{person}{Elena~L. Glassman}.}
  \bibinfo{year}{2022}\natexlab{b}.
\newblock \showarticletitle{Expectation vs. Experience: Evaluating the
  Usability of Code Generation Tools Powered by Large Language Models}. In
  \bibinfo{booktitle}{\emph{CHI Conference on Human Factors in Computing
  Systems Extended Abstracts}} (New Orleans, LA, USA)
  \emph{(\bibinfo{series}{CHI EA '22})}. \bibinfo{publisher}{Association for
  Computing Machinery}, \bibinfo{address}{New York, NY, USA}, Article
  \bibinfo{articleno}{332}, \bibinfo{numpages}{7}~pages.
\newblock
\showISBNx{9781450391566}
\urldef\tempurl%
\url{https://doi.org/10.1145/3491101.3519665}
\showDOI{\tempurl}


\bibitem[van~der Lee et~al\mbox{.}(2019)]%
        {DBLP:conf/inlg/LeeGMWK19}
\bibfield{author}{\bibinfo{person}{Chris van~der Lee}, \bibinfo{person}{Albert
  Gatt}, \bibinfo{person}{Emiel van Miltenburg}, \bibinfo{person}{Sander
  Wubben}, {and} \bibinfo{person}{Emiel Krahmer}.}
  \bibinfo{year}{2019}\natexlab{}.
\newblock \showarticletitle{Best practices for the human evaluation of
  automatically generated text}. In \bibinfo{booktitle}{\emph{Proceedings of
  the 12th International Conference on Natural Language Generation, {INLG}
  2019, Tokyo, Japan, October 29 - November 1, 2019}},
  \bibfield{editor}{\bibinfo{person}{Kees van Deemter},
  \bibinfo{person}{Chenghua Lin}, {and} \bibinfo{person}{Hiroya Takamura}}
  (Eds.). \bibinfo{publisher}{Association for Computational Linguistics},
  \bibinfo{pages}{355--368}.
\newblock
\urldef\tempurl%
\url{https://doi.org/10.18653/v1/W19-8643}
\showDOI{\tempurl}


\bibitem[Wang et~al\mbox{.}(2020)]%
        {wang2020human}
\bibfield{author}{\bibinfo{person}{Dakuo Wang}, \bibinfo{person}{Elizabeth
  Churchill}, \bibinfo{person}{Pattie Maes}, \bibinfo{person}{Xiangmin Fan},
  \bibinfo{person}{Ben Shneiderman}, \bibinfo{person}{Yuanchun Shi}, {and}
  \bibinfo{person}{Qianying Wang}.} \bibinfo{year}{2020}\natexlab{}.
\newblock \showarticletitle{From human-human collaboration to Human-AI
  collaboration: Designing AI systems that can work together with people}. In
  \bibinfo{booktitle}{\emph{Extended abstracts of the 2020 CHI conference on
  human factors in computing systems}}. \bibinfo{pages}{1--6}.
\newblock


\bibitem[Weisz et~al\mbox{.}(2021)]%
        {DBLP:conf/iui/WeiszMHRRMAT21}
\bibfield{author}{\bibinfo{person}{Justin~D. Weisz},
  \bibinfo{person}{Michael~J. Muller}, \bibinfo{person}{Stephanie Houde},
  \bibinfo{person}{John~T. Richards}, \bibinfo{person}{Steven~I. Ross},
  \bibinfo{person}{Fernando Martinez}, \bibinfo{person}{Mayank Agarwal}, {and}
  \bibinfo{person}{Kartik Talamadupula}.} \bibinfo{year}{2021}\natexlab{}.
\newblock \showarticletitle{Perfection Not Required? Human-AI Partnerships in
  Code Translation}. In \bibinfo{booktitle}{\emph{{IUI} '21: 26th International
  Conference on Intelligent User Interfaces, College Station, TX, USA, April
  13-17, 2021}}, \bibfield{editor}{\bibinfo{person}{Tracy Hammond},
  \bibinfo{person}{Katrien Verbert}, \bibinfo{person}{Dennis Parra},
  \bibinfo{person}{Bart~P. Knijnenburg}, \bibinfo{person}{John O'Donovan},
  {and} \bibinfo{person}{Paul Teale}} (Eds.). \bibinfo{publisher}{{ACM}},
  \bibinfo{pages}{402--412}.
\newblock
\urldef\tempurl%
\url{https://doi.org/10.1145/3397481.3450656}
\showDOI{\tempurl}


\bibitem[Wold et~al\mbox{.}(2001)]%
        {WOLD2001109}
\bibfield{author}{\bibinfo{person}{Svante Wold}, \bibinfo{person}{Michael
  Sjöström}, {and} \bibinfo{person}{Lennart Eriksson}.}
  \bibinfo{year}{2001}\natexlab{}.
\newblock \showarticletitle{PLS-regression: a basic tool of chemometrics}.
\newblock \bibinfo{journal}{\emph{Chemometrics and Intelligent Laboratory
  Systems}} \bibinfo{volume}{58}, \bibinfo{number}{2} (\bibinfo{year}{2001}),
  \bibinfo{pages}{109--130}.
\newblock
\showISSN{0169-7439}
\urldef\tempurl%
\url{https://doi.org/10.1016/S0169-7439(01)00155-1}
\showDOI{\tempurl}
\newblock
\shownote{PLS Methods}.


\bibitem[Xu et~al\mbox{.}(2021)]%
        {DBLP:journals/corr/abs-2101-11149}
\bibfield{author}{\bibinfo{person}{Frank~F. Xu}, \bibinfo{person}{Bogdan
  Vasilescu}, {and} \bibinfo{person}{Graham Neubig}.}
  \bibinfo{year}{2021}\natexlab{}.
\newblock \showarticletitle{In-IDE Code Generation from Natural Language:
  Promise and Challenges}.
\newblock \bibinfo{journal}{\emph{CoRR}}  \bibinfo{volume}{abs/2101.11149}
  (\bibinfo{year}{2021}).
\newblock
\showeprint[arXiv]{2101.11149}
\urldef\tempurl%
\url{https://arxiv.org/abs/2101.11149}
\showURL{%
\tempurl}


\bibitem[Zhou et~al\mbox{.}(2021)]%
        {DBLP:journals/corr/abs-2105-05991}
\bibfield{author}{\bibinfo{person}{Wen Zhou}, \bibinfo{person}{Seohyun Kim},
  \bibinfo{person}{Vijayaraghavan Murali}, {and} \bibinfo{person}{Gareth~Ari
  Aye}.} \bibinfo{year}{2021}\natexlab{}.
\newblock \showarticletitle{Improving Code Autocompletion with Transfer
  Learning}.
\newblock \bibinfo{journal}{\emph{CoRR}}  \bibinfo{volume}{abs/2105.05991}
  (\bibinfo{year}{2021}).
\newblock
\showeprint[arXiv]{2105.05991}
\urldef\tempurl%
\url{https://arxiv.org/abs/2105.05991}
\showURL{%
\tempurl}


\end{thebibliography}

\end{document}